# Hybrid Machine Learning for Scanning Near-field Optical Spectroscopy

Xinzhong Chen[1*], Ziheng Yao[1,3], Suheng Xu[2], A. S. McLeod[2], Stephanie N. Gilbert Corder[3], Yueqi Zhao[4], Makoto Tsuneto[1], Hans A. Bechtel[3], Michael C. Martin[3], G. L. Carr[5], M. M. Fogler[4], Stefan G. Stanciu[6], D. N. Basov[2], Mengkun Liu[1,5*]

[1]Department of Physics and Astronomy, Stony Brook University, Stony Brook, New York 11794, USA

[2]Department of Physics, Columbia University, New York, New York 10027, USA

[3]Advanced Light Source Division, Lawrence Berkeley National Laboratory, Berkeley, CA 94720, USA

[4]Department of Physics, University of California at San Diego, La Jolla, California 92093-0319, USA

[5]National Synchrotron Light Source II, Brookhaven National Laboratory, Upton, New York 11973, USA

[6]Center for Microscopy-Microanalysis and Information Processing, Politehnica University of Bucharest, 060042, Romania

*Corresponding authors: xinzhong.chen@stonybrook.edu, mengkun.liu@stonybrook.edu.

**The underlying physics behind an experimental observation often lacks a simple analytical description. This is especially the case for scanning probe microscopy techniques, where the interaction between the probe and the sample is nontrivial. Realistic modeling to include the details of the probe is always exponentially more difficult than its "spherical cow" counterparts. On the other hand, a well-trained artificial neural network based on real data can grasp the hidden correlation between the signal and sample properties. In this work, we show that, via a combination of model calculation and experimental data acquisition, a physics-infused hybrid neural network can predict the tip-sample interaction in the widely used scattering-type scanning near-field optical microscope. This hybrid network provides a long-sought solution for accurate extraction of material properties from tip-specific raw data. The methodology can be extended to other scanning probe microscopy techniques as well as other data-oriented physical problems in general.**

At the core of a scanning probe microscopy technique lies the unique probe-sample interaction that occurs at close proximity to the sample surface. For example, the atomic force microscope (AFM) utilizes the force-distance curve to detect various material properties at the sub-nanometer scale with a sharp tip. By modeling and manipulating the tip-sample interaction, an AFM can yield quantitative information on the surface topography and atomic forces such as atomic bonding, Coulomb force, and van der Waals forces[1]. Also relying on interrogating the sample by means of a sharp probe, scattering-type scanning near-field optical microscope (s-SNOM) is now generally acknowledged as being highly useful in extracting valuable optical information at the nanoscale, enabling fundamental discoveries in physics and materials sciences over the past years[2,3]. It is based on an AFM platform and incorporates light as a third interaction, in addition to the tip and sample. Unlike far-field optics where the interaction only occurs between light and sample, the situation is fundamentally different in s-SNOM where light, sample, and probe are equally important for the outcome of the physical process that yields the signal contrast. The situation stands the same for other tip-based optical microscopies such as tip-enhanced Raman spectroscopy[4], tip-enhanced fluorescence microscopy[5], second-harmonic generation near-field optical microscopy[6], peak-force infrared microscopy[7–9], or photo-induced force microscopy[10].

In s-SNOM, light is coupled to a metal-coated AFM tip which excites and probes the evanescent waves strongly localized near the sample surface under its apex. The electrodynamic tip-sample interaction manifests itself via the modified tip radiation that can be detected in the far-field zone[11]. Through the interferometric detection of the tip-scattered field, the local sample optical properties can be measured[12] and used to derive important physical quantities such as the complex dielectric function over a broad spectral range[13–17]. Importantly, the available resolution practically depends only on the size of the tip and the sensitivity of the detector instead of the wavelength of the incident light. Quantitative pinpointing of the nanoscale spectral fingerprints is the cornerstone for many applications such as chemical identification[18–20] and quantum material characterization[21,22]. The measurement is often achieved via nano-resolved Fourier transform infrared spectroscopy (nano-FTIR) [23,24]. To interpret the nano-FTIR spectra at a quantitative level, a working understanding of the complex tip-sample interaction in s-SNOM is necessary.

Recent efforts involving analytical modeling and numerical simulations have resulted in fruitful progress such as the finite-dipole model and the lightning-rod model[25–30]. However, an exact model fitting to the experimental data is often a formidable task due to the irregular tip geometry and the electrodynamical nature of the problem. For example, it is particularly difficult for analytical models to accurately describe the near-field signal at the highly resonating regime, where the real part of the sample dielectric function $Re(\varepsilon(\omega))$ is between 0 and ~-10. Although full-wave numerical simulations can in principle model the tip-scattering exactly, the scale mismatch between the free space wavelength and interaction region requires dense meshing, making simulations computationally expensive. In this article we use two different data-driven machine learning (ML) methods to analyze the tip-scattered signal. One method utilizes an artificial neural network (NN) to learn the correlation between the sample's dielectric function $\varepsilon$ and the experimentally measured near-field signal using a limited set of nano-spectroscopy data. The other method relies on a hybrid neural network (HNN) that is developed to combine a NN and a physically motivated analytical model (Fig.1(a)). We demonstrate that the HNN in general improves the performance of the regular NN due to the infused physics especially when the training data volume is limited in quantity and quality.

Our proposed NN and HNN methods offer at least three advantages compared to all previous approaches: 1. The details of the tip geometry are implicitly encoded in the trained network and are thus unnecessary to be explicitly modeled; 2. Prediction of the signal can be made much faster than numerical simulations where a fine meshing scheme is required;

3. The network can be continuously improved upon the availability of new training data. Given the rapidly increasing amount of data generated in the field of near-field optics, and subsequent analysis requirements, we believe that the proposed data-driven approach could play an important role in enabling next-generation s-SNOM imaging and spectroscopy that pushes near-field techniques far beyond the current state of the art in terms of speed, accessibility, and reliability.

## Results

**Training data collection.** To collect the training data, five isotropic materials with well-characterized infrared (IR) dielectric functions are prepared, namely $SrTiO_3$ (STO), $(LaAlO_3)_{0.3}(Sr_2AlTaO_6)_{0.7}$ (LSAT), $NdGaO_3$ (NGO), $SiO_2$, and $CaF_2$. The corresponding wafers have been purchased (MTI Corporation) and measured using a nano-FTIR setup. Since the IR active phonons of the samples are mostly located in the mid- to far-IR regime, a customized Cu:Ge detector is used to extend the low frequency cutoff to ~330 cm$^{-1}$, enabling the maximum bandwidth to capture as much spectral information as possible. The experimental data are collected in a commercial nano-FTIR system (Neaspec GmbH) coupled to broadband synchrotron light (Advanced Light Source, Beamline 2.4, LBNL). As schematically shown in Fig. 1(b) the light is focused on the AFM tip that is oscillating at its mechanical resonance frequency $\Omega$. The detected signal is demodulated to $n\Omega$ ($n \geq 2$) to suppress the undesired far-field background. For IR frequencies it is commonly recognized that $n = 2$ achieves a good balance between signal-noise ratio (SNR) and background suppression. Therefore, through the article we refer to the signal demodulated at the second harmonics as the near-field signal, and denote it by $S_2$. The same analysis can be applied to $S_n$ with $n > 2$ as long as the SNR permits. An asymmetric Michelson interferometer is employed to obtain the interferogram of the reference signal and the tip-scattered signal. Fourier transform of the interferogram thus gives the phase-resolved spectrum with a sub 50 nm spatial resolution. The collected nano-FTIR spectra for the chosen materials are shown in the top row of Fig. 1(c). Proper normalization is required due to the nonuniform spectral intensity of the light source, scattering efficiency of the tip, and sensitivity of the detector within the frequency range. All the experimental spectra are taken with the same tip and normalized to the spectrum of a thick gold film to yield quantitatively meaningful values. In nano-FTIR measurements phase drift between sample and the thick gold film measurements could potentially occur and lead to inaccuracy. This is prevented by measuring the gold film both before and after the sample measurement and ensuring that the two spectra are identical. Due to instrumental limitations, the samples are not characterized by conventional far-field FTIR or ellipsometry to obtain the dielectric function. Instead, the corresponding dielectric functions used in this study for method development and validation are found in the literature and shown in the bottom row of Fig. 1(c) [31–35]. The spectrum and dielectric function of gold (not shown) is also explicitly used in the training[36]. Although gold does have a dispersive dielectric function in the relevant frequency range, its value is in the order of a few thousand, being reflective enough to provide a nearly frequency-independent near-field signal.

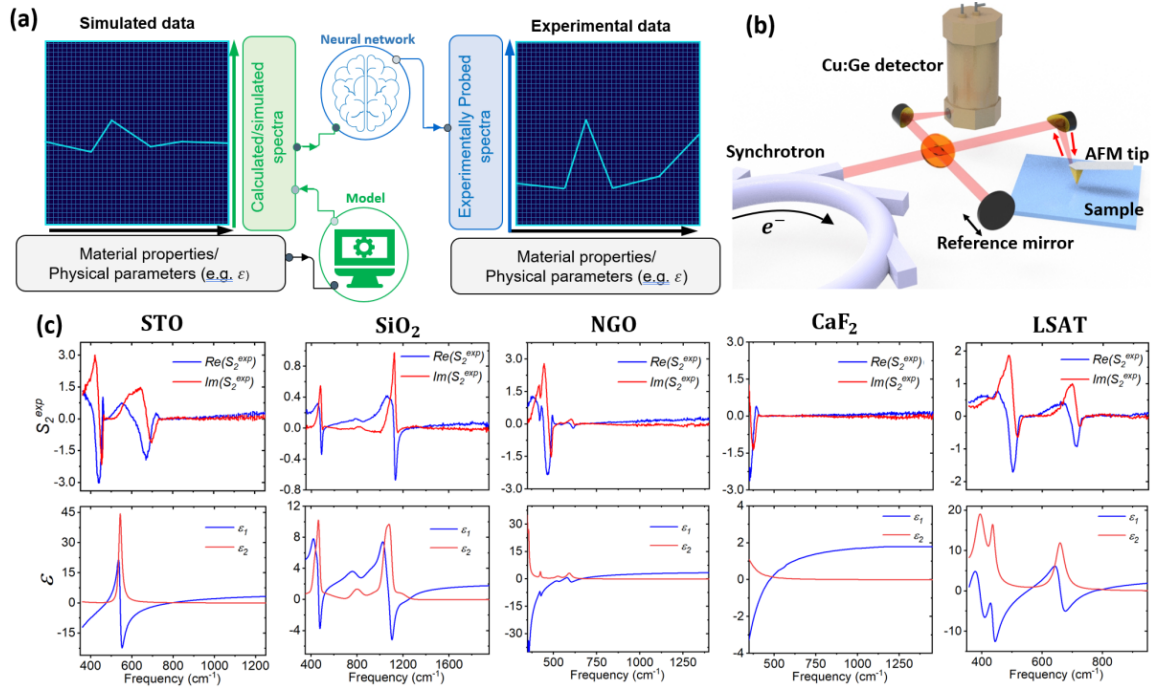

**Figure 1.** (a) Schematics of the hybrid neural network that combines a physically motivated model and a NN. The model (indicated by the computer icon) first makes a reasonable prediction based on a priori known material properties. The prediction is then fed into a NN (indicated by the brain) to be further refined based on the available training data set. (b) Schematics of the nano-FTIR setup. Broadband nano-FTIR spectra are acquired with synchrotron infrared radiation with an asymmetric Michelson interferometer setup for phase-sensitive detection. A Cu:Ge detector is used for spectral detection down to 330 cm$^{-1}$. (c) Systematically collected complex nano-FTIR spectra on five isotropic crystals (upper panels) and the corresponding dielectric functions reported in the literature (bottom panels).

We have intentionally chosen isotropic materials in this study to avoid the extra complexity caused by anisotropy. In the current body of literature it is a common practice to treat uniaxial anisotropic materials effectively as isotropic ones by using the geometric mean of the in-plane and out-of-plane dielectric functions. However, the proper treatment for biaxial material is still unclear. Nevertheless, anisotropic and layered systems / thin films are fundamentally interesting study objects for ML applications and thus our intention is to extend the proposed method towards them once sufficient training data become available. It is also well known that different tip geometries and tapping amplitudes will have a significant impact on the measurement results. To ensure consistency, a "nano-FTIR" probe (from Neaspec GmbH) has been used to collect the training data, keeping a fixed tapping amplitude of 68 nm throughout the data collection process.

**Application of NN to the training data.** Various machine learning algorithms have found their applications in a number of disciplines of academic research[37], including scanning probe microscopy[38–41]. We recently demonstrated that ML can be employed to decipher the relationship between the sample's optical constants and s-SNOM measurements with simulated data[42]. In this section we extend these efforts and realize this idea with experimental data shown in Fig 1(c).

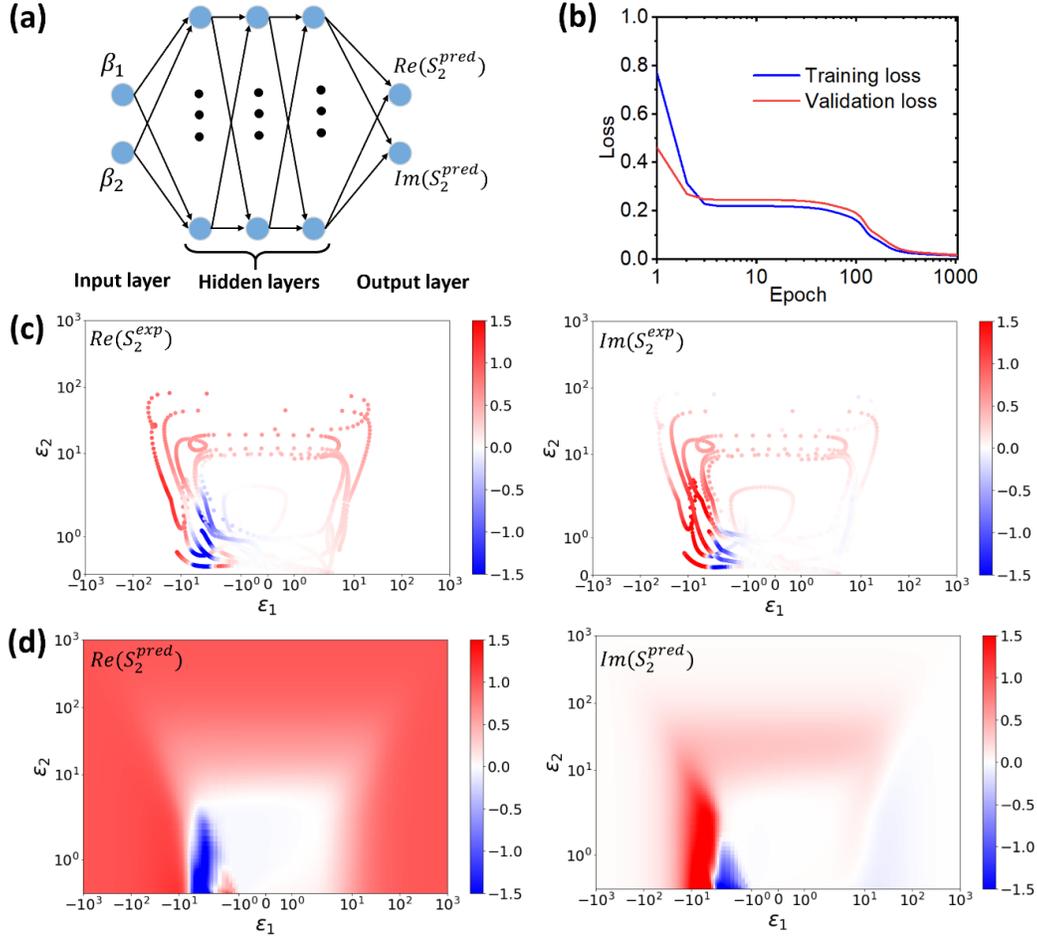

**Figure 2.** (a) Schematics of the NN with 3 hidden layers and 8 neurons per layer. (b) A typical trajectory of the training and validation losses as a function of epoch. (c) Visualization of the complex training data in the $\varepsilon$ space. (d) The predicted function of $S_2(\varepsilon)$ by the trained NN.

In this supervised learning problem, the features (input) can be the complex dielectric function $\varepsilon = \varepsilon_1 + i\varepsilon_2$ and the labels (output) the complex near-field signal $S_2^{exp} = Re(S_2^{exp}) + iIm(S_2^{exp})$ (the reverse problem will be discussed in a later section). However, a more natural representation of the optical constant is $\beta = \beta_1 + i\beta_2 = \frac{\varepsilon-1}{\varepsilon+1}$, where $\beta$ is the quasistatic reflection coefficient that encodes the sample response in the near-field interaction[26,27,43]. Without loss of generality, we use the real and imaginary parts of $\beta$ as the input features while $\varepsilon$ can be easily recovered from $\beta$ by $\varepsilon = \frac{1+\beta}{1-\beta}$. As schematically shown in Fig. 2(a), a NN with 2 inputs and 2 outputs is built using the popular ML framework TensorFlow. To avoid potential overfitting and ensure training efficiency, we do not employ a very deep network. A fully connected NN with 3 hidden layers and 8 neurons per layer is built. We find that deeper networks yield similar performance. The nonlinearity of the network is realized by using the sigmoid function as activation. Adam optimizer with a learning rate of $10^{-3}$ is used to train the NN with the goal of minimizing the mean squared error. The data is split into a training set (80%) and a validation set (20%). Although the training results fluctuate slightly due to random initial conditions, in a typical training process the training loss and the validation loss simultaneously decrease to a diminishing value, indicating the NN is not overfitting nor underfitting. The mini-batching technique is used for increased training efficiency. Consequently a low-end commercial desktop computer (CPU: Intel Core i3-10100, GPU: Nvidia GTX 1660 Super) can perform the training in the order of 5 minutes.

Before discussing the training result, it is instructional to first visualize the experimental data. The real and imaginary parts of the experimental near-field signal $S_2^{exp}$ are shown in Fig. 2(c). The graphs are replots of all the spectra data in Fig. 1(c) in a way that the position of the scatterer corresponds to the sample's $\varepsilon$ while the value of the measured near-field signal is encoded in the color. In this representation it is easy to see that near-field resonance occurs in the regime where $\varepsilon_1$ is mildly negative and $\varepsilon_2$ is small, which corresponds to the polaritonic modes in, for example, the polar crystals[44]. Given that sufficient data density is necessary to obtain reliable ML results, we have included in this study materials with distinctive IR phonon resonances. Once the NN is trained, employing it for predictions of near-field signals is computationally inexpensive. In Fig 2(d) we plot the predicted near-field response $S_2^{pred}$ by the trained NN. This can be compared to the point-dipole model and finite-dipole model (FDM) predictions (see supplemental materials). The plots show expected patterns. Near $\varepsilon = 1$, $S_2^{pred}$ is approaching zero as the sample becomes completely transparent. Strong resonance with large $S_2^{pred}$ is observed where $\varepsilon$ satisfies the aforementioned resonance condition. As $|\varepsilon|$ becomes large, $S_2^{pred}$ approaches 1.

At this point, the details of the AFM tip and the complex near-field interaction have been successfully encoded in the weights and biases of the NN without being explicitly modeled. This trained NN can be used to predict quantitative spectra for untested materials with known optical properties, or be used in the reversed way to extract the optical constant from a nano-FTIR spectral through a fitting procedure, which we discuss in detail in the later section.

**Infusing the NN with physics.** NN as a universal function approximator is undoubtedly a capable tool for finding hidden patterns in the data. However, it is well known that NN is intrinsically data-hungry. In our case, a relatively large data set covering a broad frequency range is required to yield accurate results (at least 4-5 experimentally measured broadband nano-FTIR spectra with distinctive resonance features). This is often inaccessible to researchers in the community as the availability of high-brilliance broadband light sources such as synchrotron IR radiation is very limited. To remedy the lack of data in many physics-related ML problems, NNs informed with physical principles have emerged as a promising framework[45]. For example, when the relevant physical quantity in the problem follows certain governing equations, a NN connected to an operator layer can impose restrictions on the output thus reduce the requirement for the amount of training data[46–48]. Other methods where the NN is constructed with specific physics-oriented symmetry or loss function have been explored very recently[49,50]. Moreover, a multi-fidelity approach using a combination of abundant low-fidelity data and scarce high-fidelity data was originally proposed based on Gaussian process[51]. Usually the high-fidelity data are from real experimental measurements or expensive simulations while the low-fidelity data are from simple modeling or fast simulations. This multi-fidelity ideology has recently been generalized to utilize NN and demonstrate superior performance with limited high-fidelity data compared to the regular NN[52–54]. Inspired by these ideas, in the following we apply a hybrid method that combines a physical model and the NN, namely HNN.

As schematically shown in Fig. 3(a), in HNN conventional FDM is first used to calculate the near-field response $S_2^{FDM}$ from a given set of $\varepsilon$. Then the output of the model, as well as the optical constant, is used to inform a NN with 3 hidden dense layers of 8 neurons. This NN connects the FDM prediction and the experimental data. That is, the NN aims to discover the mapping $S_2^{exp}(\varepsilon) = F(S_2^{FDM}(\varepsilon))$. Since the FDM often already gives a reasonably accurate prediction, the NN only needs to refine the result by implementing small corrections based on the training data it receives. The important advantage of this approach is that the network can yield a good prediction despite the amount of training data it is trained on. When the training data is scarce, the model prediction is closer to the original FDM as fewer modifications are made in the effort to refine it. When the training data is abundant, the NN enables and implements more corrections in areas where the FDM's accuracy struggles.

It is insightful to first briefly summarize the physics of the FDM. In the FDM the tip is approximated as an elongated spheroid[25,55], which is a significant improvement over the earlier point-dipole model[43]. Under the quasi-static approximation the effective tip polarizability is given by

$$\alpha_{eff} \propto \frac{1}{2}\frac{\beta f_0}{1-\beta f_1} + 1,$$

where $f_0$ and $f_1$ are geometric factors that depend on the tip length, apex radius, and tip-sample distance[25,55]. $\alpha_{eff}$ contains the modification of the tip's dipolar radiation property due to the near-field interaction with the sample. Furthermore, a multiplicative far-field factor $(1 + r_p)^2$ needs to be included to account for the secondary reflections off the sample surface[56], where $r_p$ is the $p$-polarized far-field Fresnel coefficient evaluated at the incident/collection angle. Finally, to compare with the experimental observable $S_n$, the demodulation is carried out as

$$S_n(\omega) \propto \int_0^{\frac{2\pi}{\Omega}}(1 + r_p)^2 \alpha_{eff}(\omega, t) e^{-in\Omega t} dt.$$

Here the tip is assumed to follow a harmonic tapping motion given by $z(t) = A(1 - \cos(\Omega t))$, where $A$ is the tapping amplitude. The efficacy of the FDM has been demonstrated numerous times in the literature[57,58]. However, there are several noteworthy limitations: the real AFM tips are not elongated spheroids; the length of the tip is often close to or even larger than the incident wavelength such that the electrodynamical retardation effect might be significant. Although these inconsistencies obviously yield differences between the FDM prediction and the experimental data, the FDM still serves as a solid starting point for the HNN. More elaborate models could in principle be implemented in future studies.

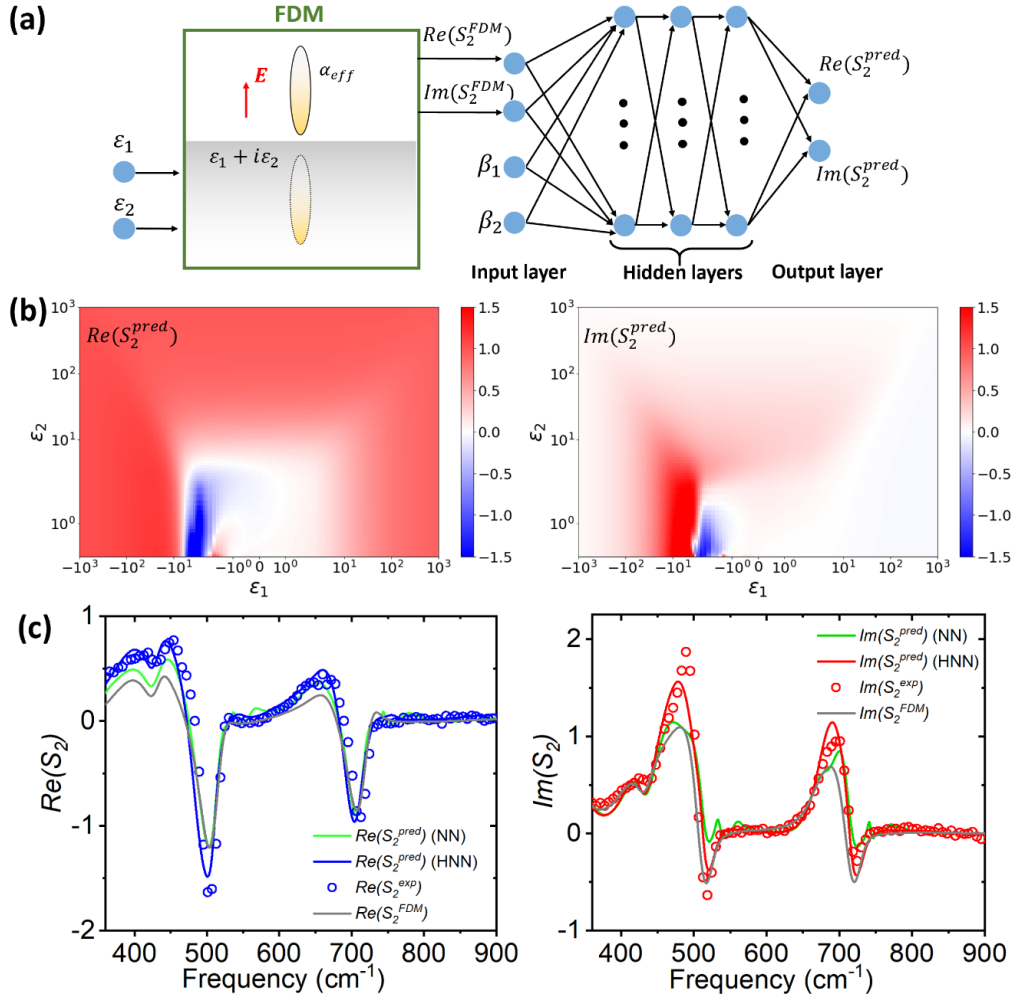

**Figure 3.** (a) Schematics of the HNN. The FDM is first used to calculate the near-field response $S_2^{FDM}$ with a given $\varepsilon$ set in the training data. The predicted $S_2^{pred}$ and $\beta$ are then used as features for the NN. The training process improves upon the original FDM prediction based on the training data. (b) The predicted function of $S_2(\varepsilon)$ by the trained HNN. (c) Comparison of the predicted $S_2$ spectra of LSAT by HNN (blue and red curves) and NN (green curves) trained without experimental LSAT data. The network predictions are also compared to experimental spectrum (circles) and FDM predictions (grey curves).

Training the HNN is up to a few times faster compared to the regular NN because the input $S_2^{FDM}$ for the dense layer is already close to the label $S_2^{exp}$. After being trained on the data in Fig. 1(c), the HNN prediction $S_2^{pred}$ is shown in Fig. 3(b). Compared to the NN prediction in Fig. 2(d), the HNN has similar features as expected but is overall smoother due to the underlying FDM. To benchmark the performance of the HNN and the regular NN, we retrain them using the original training data while keeping one spectrum out of the five as the validation set. In this specific example, the LSAT spectrum serves as validation, and the other four spectra serve as the training set for the regular NN and the HNN. After training, both networks are asked to make predictions given the LSAT dielectric function. The results are shown in Fig. 3(c). Although both the NN and the HNN are able to predict the experimental data reasonably well, HNN has a noticeably better consistency especially for the resonant peaks in the spectra. It is important to highlight that the HNN is superior not only compared to NN, but also to the original FDM, as shown in Fig. 3(d). To quantify the improvement, we calculate the total squared error (including both real and imaginary parts) for HNN prediction, NN prediction, and FDM prediction. The results are 13.27, 16.83, and 22.92, respectively. Detailed analysis of the error is presented in the Supplemental Materials. We also perform cross-validation using other material spectra as the validation set. In most cases HNN prediction is more accurate or at least similar compared to the NN counterpart.

To further validate the superiority of the HNN approach, tests using simulated data have been performed and the conclusion is in line with our observation here (see Supplemental Materials). More importantly, we also test scenarios where the training data are scarce. We show that even when the training data set only consists of one experimental spectrum, HNN can still make more accurate prediction than FDM on unseen materials while NN totally fails (see Supplemental Materials). This feature is extremely beneficial for users with limited accessibility to s-SNOM equipment or broadband light source.

**Dielectric function extraction.** Here we present our results on tackling the reverse problem – extraction of the optical dielectric function $\varepsilon$ from the experimental spectrum. In Fig. 4(a) we can visualize the training data in the $S_2$ space where the color of the scatterers represents $\varepsilon$ value. This reverse problem is of great practical interest thus has also been investigated extensively in the literature[13–17]. However, a universal approach that is quantitatively accurate has yet to be established and is by far one of the biggest obstacles in s-SNOM data analysis. This is primarily due to the inherent difficulties in accurately modeling the tip geometry and the multivalued nature of the $S_2 - \varepsilon$ mapping. In the previous study we show with simulated data that it is possible to train a NN that takes $S_2$ as the input feature and the corresponding $\varepsilon$ as the corresponding label, given that data points where $S_2$ close to unity are removed[42]. This removal process is necessary to eliminate the multi-valued issue of the reverse problem. The trained NN is able to perform prediction of $\varepsilon$ of moderately small values given the experimental spectrum $S_2^{exp}$. We test the same approach here by replacing the simulated data with experimental data but do not obtain satisfactory results. The reason is twofold. First, the noise level in the experimental data is larger than estimated in the simulated data. The size of the training data set here is also much smaller, consisting of only four spectra. So far the fitting procedure discussed above is the most reliable approach we have found.

Fortunately we can impose additional constraints, such as the continuity of the dielectric function and the Kramers-Kronig relation, to circumvent the multivalue issue. As often practiced in far-field IR spectroscopy, approximating the dielectric function as a sum of Lorentz oscillators simultaneously satisfy both requirements[59]. That is, we can write

$$\varepsilon = \varepsilon_\infty + \sum_{i=1}^{N} \frac{A_i^2}{\omega_{i0}^2 - \omega^2 - i\gamma_i \omega},$$

where $\varepsilon_\infty$ is the high frequency dielectric constant. $\omega_{i0}$, $A_i$, and $\gamma_i$ are the frequency, strength, and damping rate of the $i$-th oscillator, respectively. Since our trained NN and HNN are capable of making fast and accurate forward predictions based on $\varepsilon$, the reverse problem becomes finding the most suitable set of values for $\varepsilon_\infty$, $\omega_{i0}$, $A_i$, and $\gamma_i$ that yield the best fit for $S_2^{exp}$. To this end we first manually fit $S_2^{exp}$ by tuning $\varepsilon_\infty$, $\omega_{i0}$, $A_i$, and $\gamma_i$ while visually comparing the prediction of NN or HNN $S_2^{pred}$ to the experimental spectrum $S_2^{exp}$. Once $S_2^{pred}$ roughly recovers $S_2^{exp}$, automatic optimization algorithms such as the Broyden–Fletcher–Goldfarb–Shanno (BFGS) or Powell's method are invoked to optimize the values of $\varepsilon_\infty$, $\omega_{i0}$, $A_i$, and $\gamma_i$ until the optimal values are found via minimizing the mean squared error.

Here we demonstrate the typical results using two examples. First regular NN and HNN are trained with a training set composed of spectra of four materials, leaving out the LSAT spectrum. Then we manually choose the initial values of $\varepsilon_\infty$, $\omega_{i0}$, $A_i$, and $\gamma_i$ such that $S_2^{pred}$ roughly fit the LSAT experimental spectrum. This rough manual fitting helps avoid local minima trapping as well as optimization speed. Finally we run the BFGS algorithm to find the optimized $\varepsilon_\infty$, $\omega_{i0}$, $A_i$, and $\gamma_i$. As shown in Fig. 4(b), the HNN results in similar predictions which are very close to the nominal values from the literature. NN has a similar prediction that is practically indistinguishable from that of HNN thus is not shown. A similar procedure is carried out by replacing LSAT with STO, and the same conclusion can be drawn as shown in Fig. 4(c).

In addition, we also demonstrate an inversion attempt based purely on the multivariate optimization technique -- differential evolution[60]. This stochastic method can be used to find the $\varepsilon$ that gives the global minimal of the squared difference between the HNN prediction $S_2^{pred}$ and the experimental spectrum $S_2^{exp}$. Compared to gradient-based optimization techniques like BFGS, differential evolution can search a larger variable space thus is less likely to be trapped in a local minimum. The inversion is performed in a frequency by frequency fashion and the results are compared to Lorentzian fitting in Fig. 4(b) and 4(c). It is clearly shown that this direct inversion can give a reasonable result at most of the spectral range but fails when the value of the dielectric function is large. It is also more sensitive to noise, which limits its application in practice.

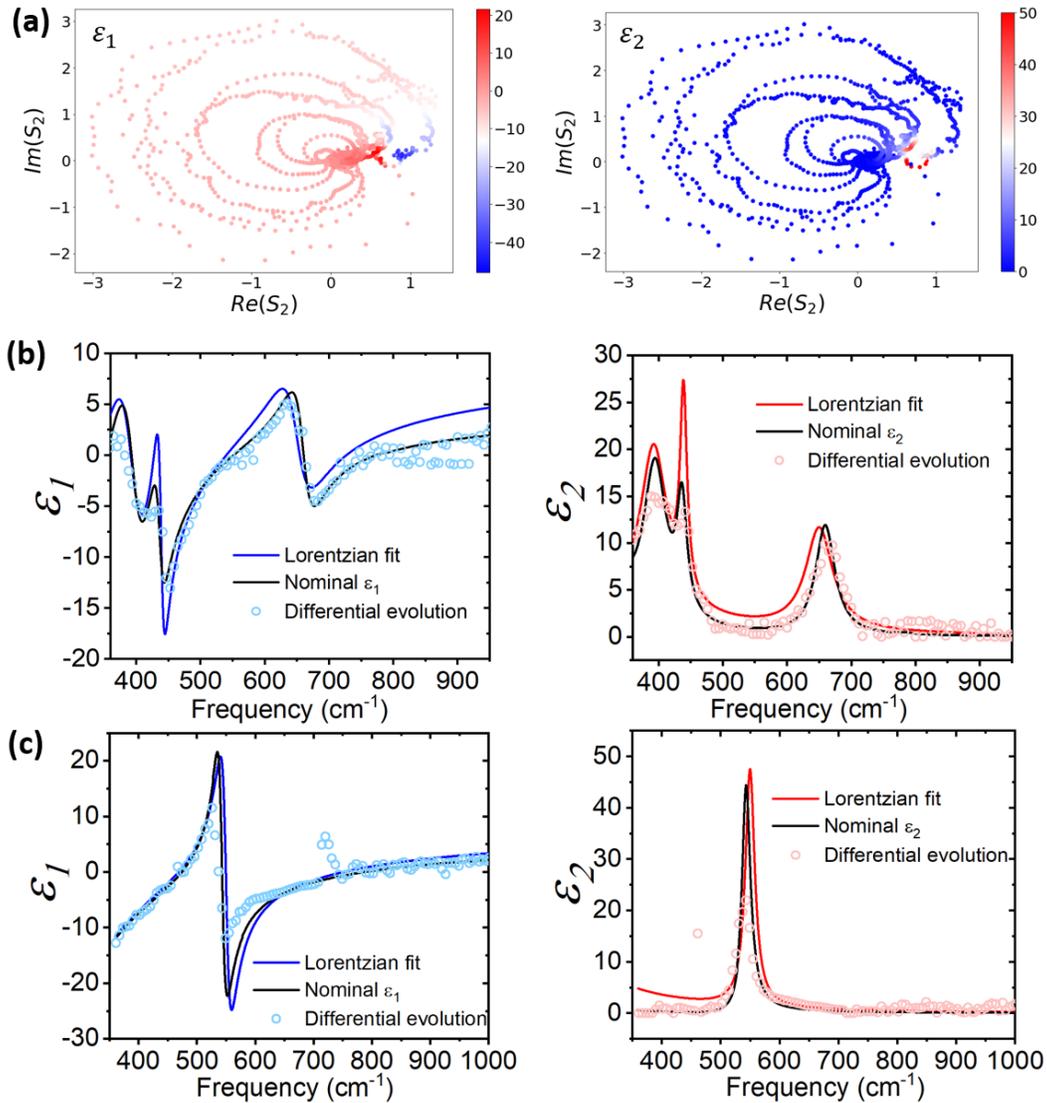

**Figure 4.** (a) Visualization of the complex training data in the $S_2$ space. (b) Comparison of the nominal dielectric function (dashed curves), the Lorentzian fitting result (solid curves), and the differential evolution result (circles) of the LSAT dielectric function (c) Comparison of the nominal dielectric function (black curves), the Lorentzian fitting result (blue and red curves), and the differential evolution result (light blue and red circles) of STO dielectric function. Differential evolution and Lorentzian fitting results are obtained using the HNN trained without the predicted material in the training set. Results from the NN are similar to those from the HNN thus not shown.

## Discussions

The fields of Machine Learning and Artificial Intelligence have expanded exponentially over the past years, harboring a large variety of techniques that can hugely augment the outputs of well-established and emerging cutting-edge imaging technologies. The translation of such approaches into concrete applications oriented towards materials or life sciences is likely to completely revolutionize these fields over the next years, given that they enable easy access to data that is experimentally difficult and expensive, or even impossible, to collect[61,62]. Although such methods are rapidly becoming acknowledged as solutions for resolving bottlenecks that have impeded specific advances and applications in bioimaging for many years, their penetration in materials sciences still lags behind. With the current work, we contribute to this situation by introducing a novel method that merges ML and physics concepts to augment and facilitate the use s-SNOM, an emerging technique that has enabled a wide palette of valuable discoveries over the past decade. The proposed HNN architecture is capable of accurately predicting the tip-sample interaction in s-SNOM, representing a long-sought solution for the highly accurate and practical extraction of material properties from tip-specific near-field data. Although the method introduced here is developed in connection to s-SNOM, it has very important value to enrich our understanding of the near-field tip-sample interactions, which are exploited in many important techniques for extracting optical information at the nanoscale[4,9,10]. Furthermore, the presented framework can be used as a backbone, in developing applications that address similar problems associated with such imaging modalities, especially in material and bio-systems.

For implementing the proposed methodology, we systematically collected a set of high-quality training data which enabled achieving the quantitative modeling of the near-field tip-sample interaction with unprecedented accuracy in a data-driven fashion. This circumvents a cumbersome and computationally challenging problem associated with past efforts that relied only on analytical modeling or numerical simulations, where detailed information of the tip has profound effects on its electrodynamic response. In the approach introduced here, explicit information about the tip is not required as it has been implicitly encoded into the training data and thus is intrinsically "learned" by the proposed ML algorithm.

We demonstrate reversible forecast of s-SNOM signals and specific material properties with our approach. In particular, we show that both the 'material property → near-field signal' and 'experimental near-field data → material property' translation can be achieved with superb accuracy and efficiency. The importance of the latter is of considerable value to facilitate further s-SNOM applications, as many times the raw datasets collected with this technique are difficult to interpret. This prohibits its wider use and biases its popularity, keeping s-SNOM applications confined to a limited number of highly experienced s-SNOM practitioners.

An important finding of our work is that, due to the inclusion of the physically motivated model, the HNN yields quantitatively more consistent predictions than the NN architecture. A major advantage of the HNN method is its flexibility, in the sense that it can accommodate both sparse or abundant training sets. The latter case is favorable for more accurately refining the simulated results, but the method is not restricted to specific training data volume. The architecture of the proposed HNN can easily accommodate the inclusion of additional training sets once these become available.

It is important to realize that there are still limitations in our current study. First of all, we restrict ourselves to the simple relation between $S_2$ and $\varepsilon$, while leaving other parameters such as tapping amplitude, tip apex radius, and sample thickness or anisotropy unexplored. Some of these parameters are especially important in the strong resonance regime. For example, differences in tapping amplitude, tip-sample distance, or tip geometry can result in significantly different spectra (see Supplemental Materials)[9,63]. Those parameters can be further investigated in future studies with dedicated training data sets. In addition, we ignore the explicit wavelength dependence, assuming $S_2$ is purely determined by $\varepsilon$. Experimental evidence has shown that this is a good approximation but $S_2$ does have a subtle wavelength dependence, likely due to the discrete antenna resonance modes of the tip (see Supplemental Materials). This is especially relevant at terahertz frequency range[64]. Furthermore, although in this work we chose well-documented materials as training samples, uncertainties in the dielectric functions do exist as the fabrication and growth conditions of the crystals vary. For optimal results, the dielectric functions of the samples should in principle be individually characterized.

Besides investigating the correlation among material properties, experimental parameters, and measurement readouts, other aspects of the near-field optics can benefit from ML techniques as well. For example, extracting the polariton wavelength and quality factor from near-field images of various van der Waals materials is routinely done by manual fitting procedures[65,66]. A well-trained convolutional NN could potentially automate this process and obtain accurate results with much higher efficiency. Another interesting direction is the ML-based super-resolution, which has been successfully applied to different optical microscopy techniques [61,62,67,68]. Better resolution can in principle be achieved in near-field imaging using ML-assisted algorithms, which would help circumvent current problems such as the fabrication of ultra-sharp tips, limited detector sensitivity, and insufficient signal-to-noise ratio. Other applications such as optimizing the tip design for enhanced scattering can benefit from the help of ML as well[69,70].

In conclusion, we have demonstrated that the HNN is a highly capable tool for learning the hidden patterns in the near-field response in s-SNOM, and that it provides unmatched results in terms of extracting very specific and important material properties from the raw s-SNOM datasets. In applications that require it, (e.g. s-SNOM system testing, s-SNOM training data set generation, etc.) such forecast can also be done in reverse. We hope that our results will encourage and serve as a good starting point for future explorations of ML on nano-optics, and to facilitate such efforts our trained network is attached as supplemental materials, together with a link to the datasets used in this experiment for training.

## Acknowledgments

X.Z.C, M.K.Liu, and D.N.Basov acknowledge support from the U.S. Department of Energy, Office of Science, National Quantum Information Science Research Centers, Co-design Center for Quantum Advantage ($C^2QA$) under contract number DE-SC0012704. S.G.S. acknowledges the support of UEFISCDI grant RO-NO-2019-0601 (MEDYCONAI). This research used resources of the Advanced Light Source, a U.S. DOE Office of Science User Facility under contract no. DE-AC02-05CH11231. Z.Y. acknowledges partial support from the ALS Doctoral Fellowship in Residence Program.

# Supplemental materials for

# Hybrid Machine Learning for Scanning Near-field Optical Spectroscopy

Xinzhong Chen[1*], Ziheng Yao[1,3], Suheng Xu[2], A. S. McLeod[2], Stephanie N. Gilbert Corder[3], Yueqi Zhao[4], Makoto Tsuneto[1], Hans A. Bechtel[3], Michael C. Martin[3], G. L. Carr[5], M. M. Fogler[4], Stefan G. Stanciu[6], D. N. Basov[2], Mengkun Liu[1,5*]

[1]Department of Physics and Astronomy, Stony Brook University, Stony Brook, New York 11794, USA

[2]Department of Physics, Columbia University, New York, New York 10027, USA

[3]Advanced Light Source Division, Lawrence Berkeley National Laboratory, Berkeley, CA 94720, USA

[4]Department of Physics, University of California at San Diego, La Jolla, California 92093-0319, USA

[5]National Synchrotron Light Source II, Brookhaven National Laboratory, Upton, New York 11973, USA

[6]Center for Microscopy-Microanalysis and Information Processing, Politehnica University of Bucharest, 060042, Romania

*Corresponding authors: xinzhong.chen@stonybrook.edu, mengkun.liu@stonybrook.edu.

1. **Demonstrating the validity of the HNN using simulated data**

In the main text we have shown that the HNN generalizes well to unseen data. However, due to the limited amount of data we have, it is difficult to perform further validation. Here to demonstrate the validity and superiority of the proposed HNN. We perform multiple tests on simulated data where the ground truth is known.

As a simple start, we attempt to realize the two examples shown by Meng and Karniadakis[1]. In the first example, the single variate function we try to recover within the [0,1] interval has the form $y_h = (6x - 2)^2 \sin(12x - 4)$ (red curve in Fig. S1(a)). We pretend this function is unknown and only sampled at four points as indicated by the red circles (high-fidelity data). In addition, we have another correlated function whose form is known and densely sampled as $y_l = 0.5(6x - 2)^2 \sin(12x - 4) + 10(x - 0.5) - 5$ (low fidelity data, blue curve Fig. S1(a)). We set up a HNN whose inputs are both $x$ and $y_l(x)$ with the label being $y_h(x)$. After trained with only four high-fidelity data points, the HNN prediction of $y_h$ in the [0,1] interval is shown as the red dots in Fig. S1(a), which is fully consistent with the functional form of $y_h$. Another example with $y_h = (x - \sqrt{2})\sin(8\pi x)^2$ and $y_l(x) = \sin(8\pi x)$ is shown in Fig. S1(b).

This approach can be naturally generalized to multivariate functions. In the next test we try to recover $f_h(x, y) = x^2 \sin(y) + xy$ but all we know is 500 high-fidelity data randomly sampled on $x$-$y$ plane and the low-fidelity correlated function $f_l(x, y) = 3x^2 \sin(y) + 4xy$. We train a HNN (with the data and the correlated function) as well as a NN (with only the data). The results are shown in Fig. S1(c). Although predictions $f^{pred}$ from both HNN and

NN recover the features of $f_h$, HNN does a better job at avoiding artefacts due to under sampling.

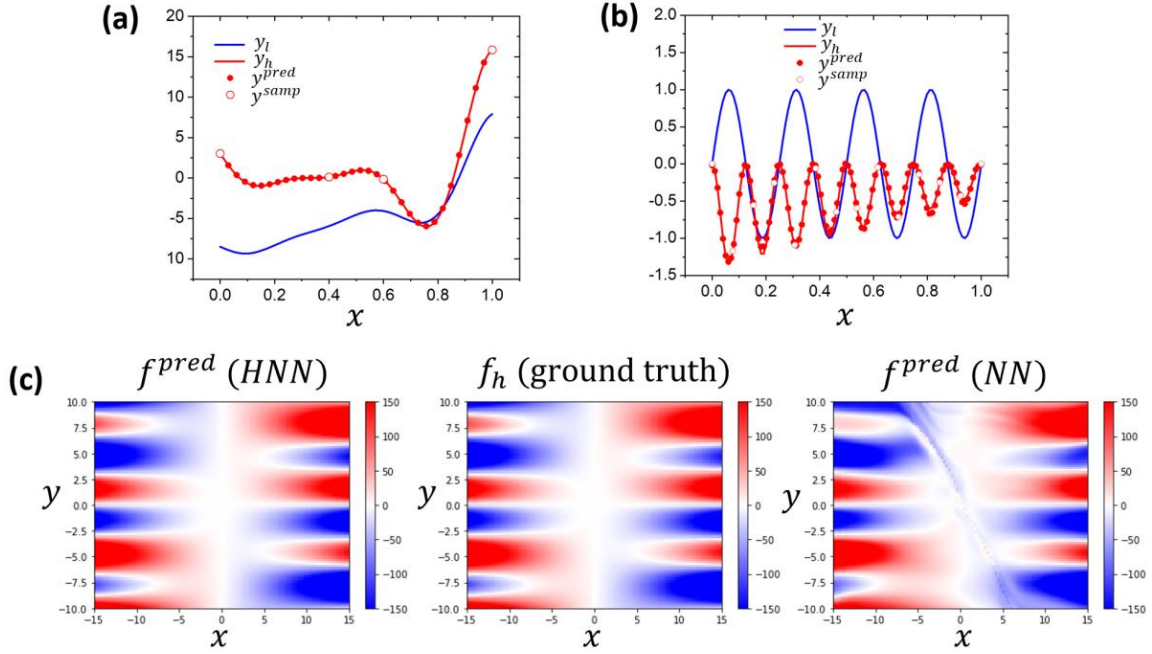

**Figure S1.** (a) and (b) Examples of HNN on single variate functions; (c) Example of HNN and NN on multivariate variate function.

To mimic and verify the approach demonstrated in the main text, in the next test we pretend the finite dipole model (FDM) is unknown and aim to recover the FDM prediction $S_2^{FDM}(\varepsilon)$ with the help of a known but less rigorous point-dipole model (PDM). We stress that this is completely analogous to what we have done in the main text – recovering the unknown experimental relation $S_2^{exp}(\varepsilon)$ with the help of the known FDM.

Training data are generated based on the same materials shown in Fig. 1(c). The dielectric functions of the five materials, as well as gold, are fed into the FDM to calculate the corresponding spectra. The result is shown in Fig. S2(a). Next, we use the HNN to recover $S_2^{FDM}(\varepsilon)$ only using this training data set and the PDM, as schematically shown in Fig. S2(b). In the PDM the tip is approximated as a sphere with radius $a$. The effective polarizability of the sphere above the sample surface is given by $\alpha_{eff} = \dfrac{\alpha}{1 - \dfrac{\alpha\beta}{16\pi(z+a)^3}}$, where $\alpha = 4\pi a^3$ is the bare sphere polarizability. Following the same training procedure as in the main text, the trained HNN is able to make fast predictions on a given $\varepsilon$. The HNN predicted $S_2^{pred}(\varepsilon)$ is shown in Fig. S2(c). In Fig. S2(d) we plot the $S_2^{FDM}(\varepsilon)$ using the analytical formula discussed in the main text with same parameters in producing Fig. S2(a), serving as the ground truth. In the comparison, it is clear that HNN successfully recover $S_2^{FDM}(\varepsilon)$ with only very minor inconsistency barely noticeable to human eyes. To further show the superiority of HNN compared to NN, training data in Fig. S1(a) is used to train a NN with the same architecture as the one used in the main text. $S_2^{pred}(\varepsilon)$ from this NN is shown in Fig. S2(e). Compared to the ground truth, although most features are accurately recovered, in certain regimes the deviations are evident.

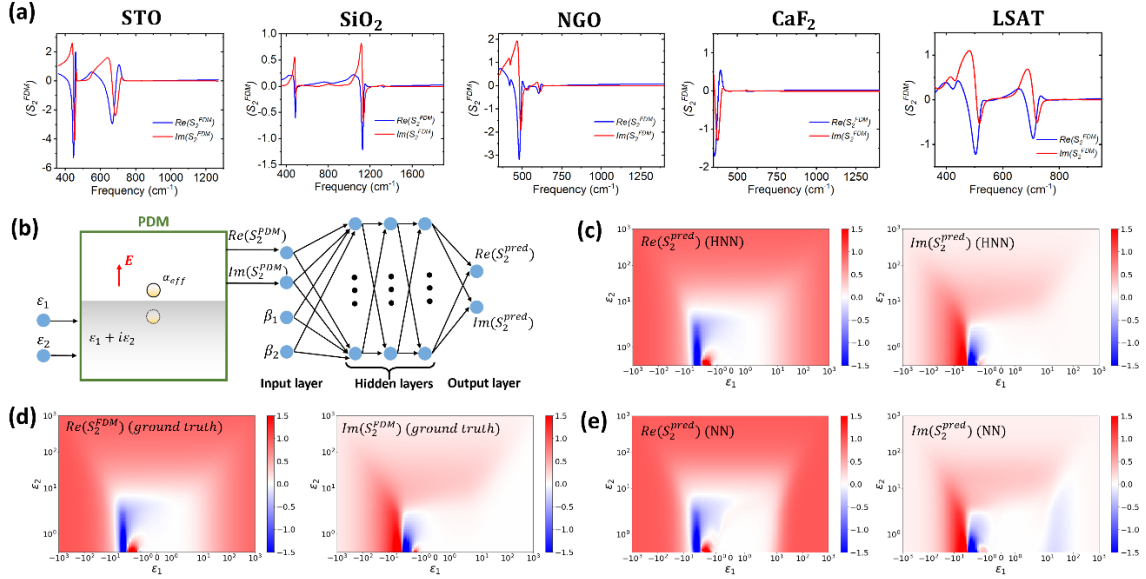

**Figure S2.** (a) Training data on the five materials. The spectra are simulated from the FDM. (b) Schematics of the HNN, where the PDM is used to help the training. (c) $S_2^{pred}$ from the trained HNN. (d) Analytically calculated $S_2^{FDM}$, serving as the ground truth to benchmark the prediction of HNN and NN. (e) $S_2^{pred}$ from the trained NN.

## 2. Superiority of HNN when data are scarce

Here we demonstrate the superiority of HNN compared to NN when the data are scarce. In this example, assume the only training data we have is one spectrum measured on NGO. A HNN and a NN with the same architecture as shown in the main text are trained on the NGO spectrum then used to make prediction on, for example, LSAT spectrum. The results are shown in Fig. S3.

Clearly NN fails to make a reasonable prediction. This should be expected as without additional physical information the NN requires a much larger training data volume to establish the relation between input features and the labels. HNN on the other hand is able to fit the experimental spectrum well, also offering an overall improvement over the FDM. This example demonstrates the power of HNN and the benefit of including physics into the network.

## 3. Potential inconsistency in the training data and future improvements

High quality data is of paramount importance for a ML project. Inconsistency in the training data could confuse the ML models thus potentially result in wrong conclusions. Here we discuss potential sources of noises and inconsistencies in our training data. We do realize that there is minor imperfection in our training data but the general conclusions we draw are not affected.

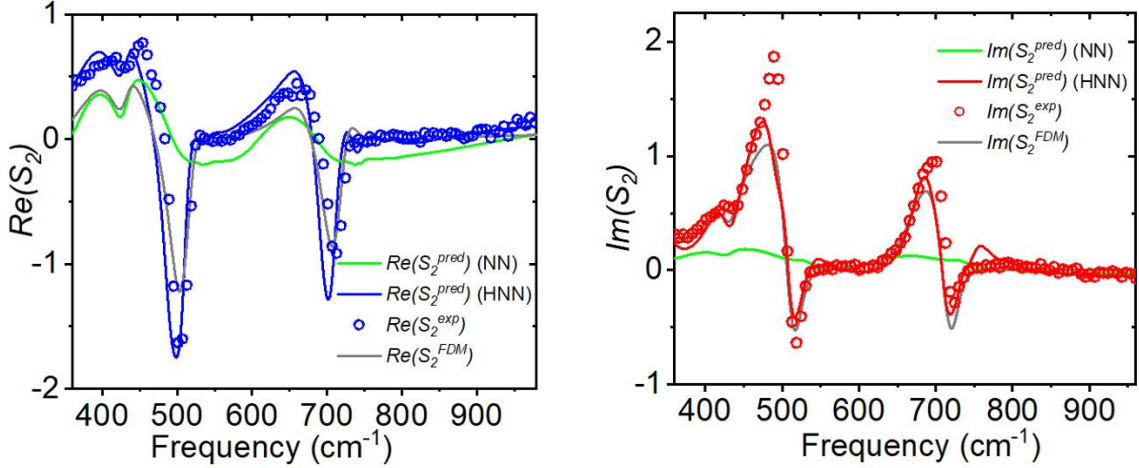

**Figure S3.** HNN and NN predictions on LSAT after trained only on NGO spectrum. The predictions are compared to experimental spectrum (ground truth) and FDM calculation.

### i. Frequency-dependent near-field response

Our study is based on the assumption that the near-field response is frequency-independent. This is, to a certain degree, a good approximation. However, we do observe a mild frequency-dependence in the data. In Fig. S4 we show a $S_2$ spectrum measured on an intrinsic Si wafer. The spectrum is normalized to a thick gold film, the same as other training data in Fig. 1(c). Although Si does not have a dielectric function dispersion in the near- to mid-IR regime, the spectrum is not homogeneously flat. This spectral variation is highly repeatable and above the noise threshold. Since the wavelength of the incident light is in the same order of magnitude as the length of the tip, we attribute it to the antenna response of the tip. This frequency-dependence of the near-field response still awaits future research. It could in principle be modeled using numerical simulation and incorporated into the network.

### ii. Spectral variation due to small change of the experimental parameters

Throughout the data acquisition, we try to keep the experimental parameters such as tip-tapping amplitude constant. However, retracting and approaching the tip during sample exchanges could lead to small fluctuation in the experimental parameters as well as the optical alignment. The fluctuation, a few nm in tapping amplitude, for example, is not observable in the collected spectra in most cases, but when strong resonance is present this small change could be amplified. It has been shown that the near-field response is sometimes not a monotonic function of the tip-sample distance in the strong resonance regime[2,3]. Therefore, the minimal tip-sample distance and tapping amplitude become crucial factors for the final spectral shape. This could potentially lead to inconsistency in the training data. Future studies with larger data volume should average out the inconsistency.

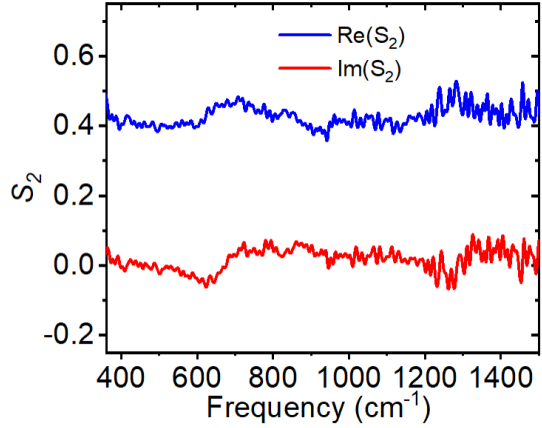

**Figure S4.** $S_2$ spectrum taken on an intrinsic Si wafer and normalized to a thick gold film. Repeatable spectral variation beyond the noise threshold is clearly observed.

iii.     **Variation in sample dielectric function**

The dielectric functions of the chosen materials are extracted from the literature. It is often the case that different groups report slightly different dielectric functions for the same material. This is either due to the margin of error in characterization or variations in sample quality and fabrication/growth conditions. To maximize consistency, individual samples used for ML training should be characterized via far-field FTIR or ellipsometry. Unfortunately due to instrumental limitations, it is not realized in the current study but can be implemented in future studies.

4. **Prediction error around resonance regime**

As discussed above, near the resonance regime the measurement is very sensitive to external experimental parameters. The geometry of the tip is also a crucial factor for the near-field response we experimentally measure. This is why analytical models using unrealistic tip geometries often have relatively poor predictability for strong resonance. Here we analyze the error of the HNN prediction and the FDM prediction. In the left panel of Fig. S5(a) we show the squared error $\left(S_2^{pred} - S_2^{exp}\right)^2$ of HNN prediction (blue curve in Fig. 3(c)) and the experimental spectrum on LSAT (blue circles in Fig. 3(c)). Significant error only occurs around the resonance peak where the dielectric function is close to -5 with a small imaginary part. In the right panel of Fig. S5(a) we show the squared error $\left(S_2^{FDM} - S_2^{exp}\right)^2$ of FDM prediction (grey curve in Fig. 3(c)) and the experimental spectrum on LSAT (blue circles in Fig. 3(c)). Clearly a larger overall error in $\left(S_2^{FDM} - S_2^{exp}\right)^2$ is observed, which indicates the improvement of the HNN over the FDM. The largest error also occurs around the resonance.

We further explore the discrepancy between the HNN prediction and the FDM prediction. The real and imaginary parts of $S_2^{pred} - S_2^{FDM}$ are shown in Fig. S5(b). Again, the discrepancy clearly is the largest around the resonance regime. How to better capture the strong resonance behavior is still an ongoing quest, which we hope to resolve when a larger training data volume is available in the near future.

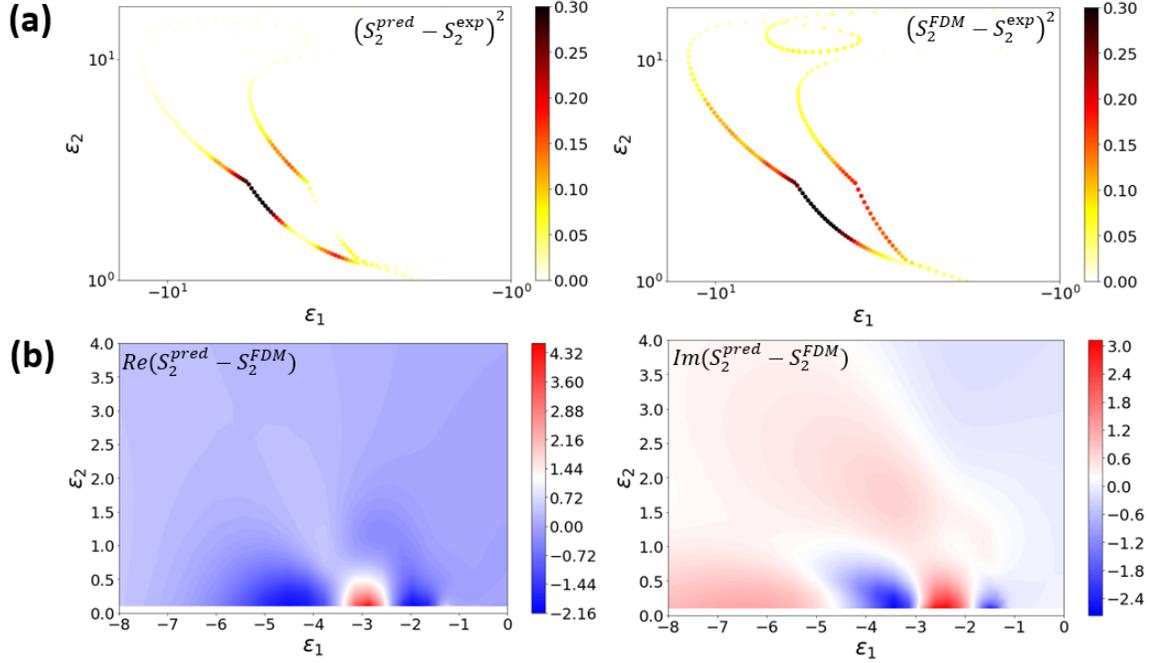

**Figure S5.** (a) Errors of HNN prediction (left) and FDM prediction (right) for LSAT spectrum as a function of dielectric function. (b) Difference in HNN prediction and FDM prediction as a function of dielectric function.

## 5. Other ML algorithms

NN-based algorithms have been proven over and over again to be a versatile tool for modeling the data. Therefore, they make up a sizable portion of the ML algorithms. It is important to note that in the broad field of ML there exists a wide range of algorithms. In our previous study we briefly surveyed a few of the popular ones such as kNN and decision tree[4]. Here we again experiment with these algorithms using experimental data and the results are shown in Fig. S6.

So far these algorithms do not seem to perform better than NN-based models. However, more sophisticated algorithms such as multi-fidelity Gaussian process await to be explored in future studies.

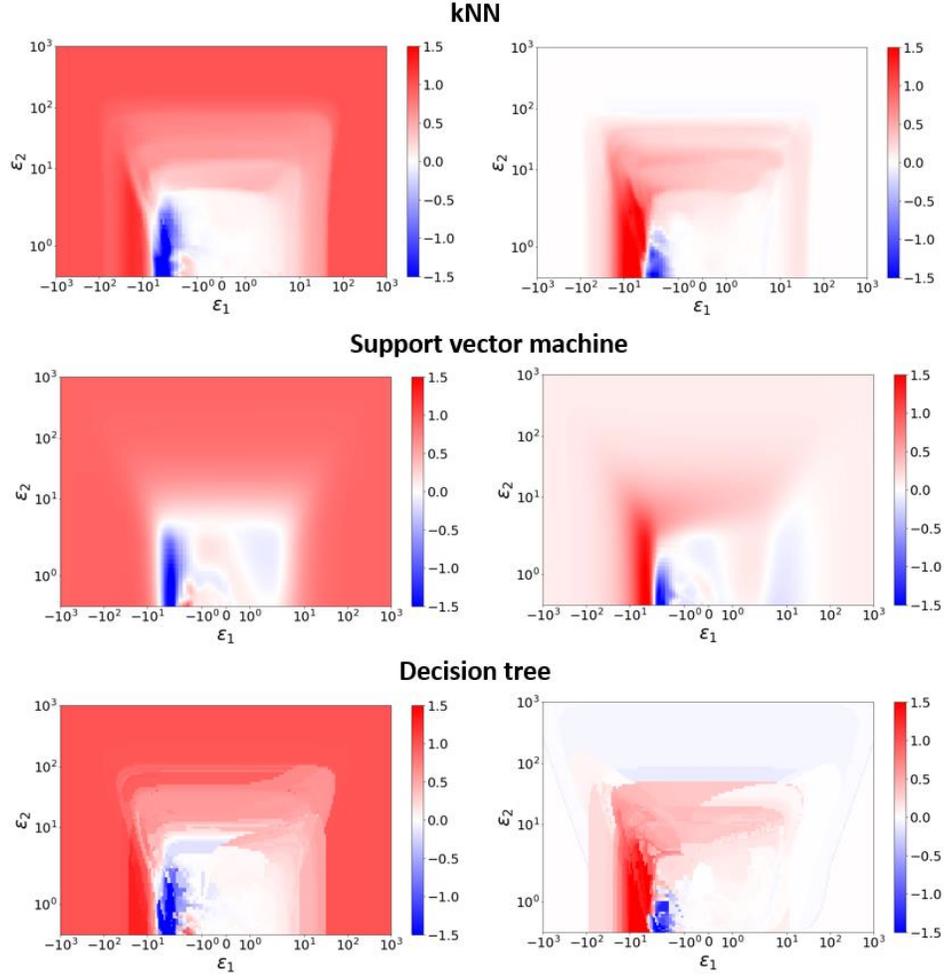

**Figure S6.** $Re(S_2^{pred}(\varepsilon))$ and $Im(S_2^{pred}(\varepsilon))$ given by kNN, support vector machine, and decision tree.